\documentclass[aps,prl,groupaddress,twocolumn,floatfix]{revtex4-2}
\usepackage{units}
\usepackage{amsmath}
\usepackage{amssymb}
\usepackage{graphicx}
\usepackage{bm}
\usepackage{multirow,xcolor,relsize,ulem,microtype}

\newcommand{\be}{\begin{equation}}
\newcommand{\ee}{\end{equation}}

\newcommand{\re}[1]{\text{Re}[#1]}
\newcommand{\im}[1]{\text{Im}[#1]}

\newcommand{\ccc}[1]{{\color{black}#1}}

\makeatletter
\renewcommand*\env@matrix[1][*\c@MaxMatrixCols c]{%
  \hskip -\arraycolsep
  \let\@ifnextchar\new@ifnextchar
  \array{#1}}
\makeatother

\begin{document}

\title{Linear localization of zero modes in weakly coupled non-Hermitian reservoirs}

\author{Bingkun Qi}
\affiliation{\textls[-18]{Department of Physics and Engineering, College of Staten Island, CUNY, Staten Island, NY 10314, USA}}
\affiliation{The Graduate Center, CUNY, New York, NY 10016, USA}

\author{Li Ge}
\email{li.ge@csi.cuny.edu}
\affiliation{\textls[-18]{Department of Physics and Engineering, College of Staten Island, CUNY, Staten Island, NY 10314, USA}}
\affiliation{The Graduate Center, CUNY, New York, NY 10016, USA}

\date{\today}

\begin{abstract}
Topological and symmetry-protected non-Hermitian zero modes have attracted considerable interest in the past few years. Here we reveal that they can exhibit an unusual behavior when transitioning between the extended and localized regimes: When weakly coupled to a non-Hermitian reservoir, such a zero mode displays a linearly decreasing amplitude as a function of space, which is not caused by an EP of a Hamiltonian, either of the entire system or the reservoir itself. Instead, we attribute it to the non-Bloch solution of a linear homogeneous recurrence relation, together with the underlying non-Hermitian particle-hole symmetry and the zeroness of its energy. 
\end{abstract}

\maketitle

\section*{Introduction}

Wave localization is one of the most celebrated physical phenomena of the past century \cite{Localization_RMP2,Localization_RMP1,Localization_PT,Kramer}. The band picture emerged in the 1920s successfully explained localization of noninteracting or linear waves in a periodic structure \cite{Bloch}, which revealed the distinction between normal metals and insulators.
The existence of a sizable bandgap is also crucial for the robustness of topological insulators and Majorana zero modes \cite{Hasan,Qi,Alicea,Beenakker,Sarma_RMP}. Take the Su-Schrieffer-Heeger (SSH) Hamiltonian \cite{SSH} for example, which is the simplest model that offers insight into the topological origin of edge states. As the difference between the two alternate nearest-neighbor couplings reduces, so does the bandgap between its two bands that leads to the divergence of   
the localization length of its zero mode(s). 

A similar transition from localized states to extended states takes place in disordered systems \cite{Localization_RMP1,Localization_PT}. Again, by neglecting interaction and nonlinear effects, Philip Anderson pointed out that disorder can completely suppress diffusion, and all eigenstates of the system become exponentially localized \cite{Anderson}. Nevill Mott also noted that in three dimensions, there exists a mobility edge that separates localized modes and extended modes in terms of their energies \cite{Mott}. However, unlike the straightforward transition mentioned for the SSH model above, the wave functions at the mobility edge display interesting features such as power law localization \cite{Varga} and fractal dimensionality \cite{Aoki}. 

Recently, wave localization behaviors in non-Hermitian systems \cite{RMP,NPreview,NPhyreview,NMatreview} also attracted considerable attention, especially due to the non-Hermitian skin effect \cite{Hatano,Longhi_gauge,Li_gauge,Song_gauge,Szameit_gauge,Zhang_gauge,Wang_Nature}. In the presence of a spatially uniform non-Hermitian gauge field, a significant percentage of modes in the system are localized toward one edge of the system, which holds in both one-dimensional (1D) and higher-dimensional systems. By varying this imaginary gauge field spatially, one can also achieve localization at any target position \cite{Ge_gauge}, including one corner, all corners, or any interior point, as recently demonstrated using a two-dimensional array of optical micro-ring resonators \cite{Feng_gauge,Feng_gauge2}. The non-Hermitian skin effect, nevertheless, does not display a nontrivial transition to delocalization: It is exponential in nature, and it vanishes together with the imaginary gauge field, returning the system back to its underlying Hermitian properties. 

Another noteworthy developement in non-Hermitian systems involved topological zero modes. A zero mode, as the name suggests, is a zero-energy state that is often the result of symmetry protection. This zero energy in low-energy physics is typically set to a well-defined level, such as the Fermi level in condensed matter systems, or in the case of systems with coupled elements, the energy of a particular resonance of interest in one element. Different from their Hermitian counterparts, these non-Hermitian zero modes can live on the imaginary axis of the complex energy plane and be attributed to either non-Hermitian particle-hole (NHPH) symmetry \cite{NHFlatband_PRL,NHFlatband_PRJ,kawabata,Okugawa,Wu}, anti-PT symmetry \cite{antiPT1,antiPT2}, or anti-pseudo-Hermiticity \cite{Scolarici}, whether or not the bulk is periodic in construction \cite{zeromodeLaser}. 
Recent experiments in photonics \cite{Baboux,Poli,St-Jean,Bandres,Zhao,Pan,Parto} and other related fields demonstrated such exotic states using spatially arranged non-Hermitian elements, using for example, evanescently coupled optical waveguides \cite{Guo,Ruter,Heinrich} and microcavities \cite{Peng,Chang,Nature_EPsensing2}. 
The localization property of these non-Hermitian zero modes inside the band gap of the hosting bulk is well known, but related studies in a non-Hermitian reservoir has largely been lacking, especially when the reservoir does not have a localized mode itself. 

In this work, we probe whether a symmetry and topologically protected zero mode exhibits exotic behaviors in a non-Hermitian reservoir.
Remarkably, we observe a ``linear localization" phenomenon when a zero mode transitions between localized and extended regimes in the reservoir: 
If weakly coupled to this reservoir, modeled by a lattice with gain and loss modulation, a zero mode can display a linearly decreasing amplitude as a function of space in the reservoir. 
This linear tail is clearly not the diminishing localization limit of an exponential function; a vanishing exponent in $|\psi(x)|\propto e^{-x/\xi}$ indicates an infinite localization length $\xi$ and a uniform amplitude in space. Furthermore, a power law decay $|\psi(x)|\propto x^{-\beta}$ with $\beta<1/2$ is often discarded in the study of wave localization, because such a wave function cannot be normalized \cite{algebraic}. However, as we will show below, the wave function with linear localization terminates at the far end of the finite-sized non-Hermitian reservoir, and hence the lack of normalizability does not prevent it from being physical and observable. We also stress that linear localization we identify occurs \textit{away} from exceptional points (EPs) \cite{EP} of the entire lattice. It is also manifested in an \textit{eigenstate} of the whole system, unlike the polynomial behavior exhibited by a Jordan chain vector in the asymptotic limit of a long time or spatial propagation \cite{LinearTemporal,Mei,NHFlatband_PRJ}.

Below we first discuss briefly NHPH symmetry and the resulting non-Hermitian zero modes in our combined lattice. We then apply a first-order perturbation theory to verify the linearly localized zero mode inside the reservoir. By utilizing the zero-energy property, we further derive a linear homogeneous recurrence relation
\be
\Psi_n = 2\alpha \Psi_{n-2} - \Psi_{n-4}\label{eq:iter}
\ee
for the zero-mode wave function inside the reservoir, where $\alpha\in\mathbb{R}$ is a constant independent of the lattice site index $n$. Despite the periodic construction of the reservoir, the characteristic equation of this recurrence relation allows a non-Bloch solution when it has a single root of multiplicity 2, which we show is responsible for linear localization. 


\section*{Results}

\begin{figure}[t]
\centering
\includegraphics[width=\linewidth]{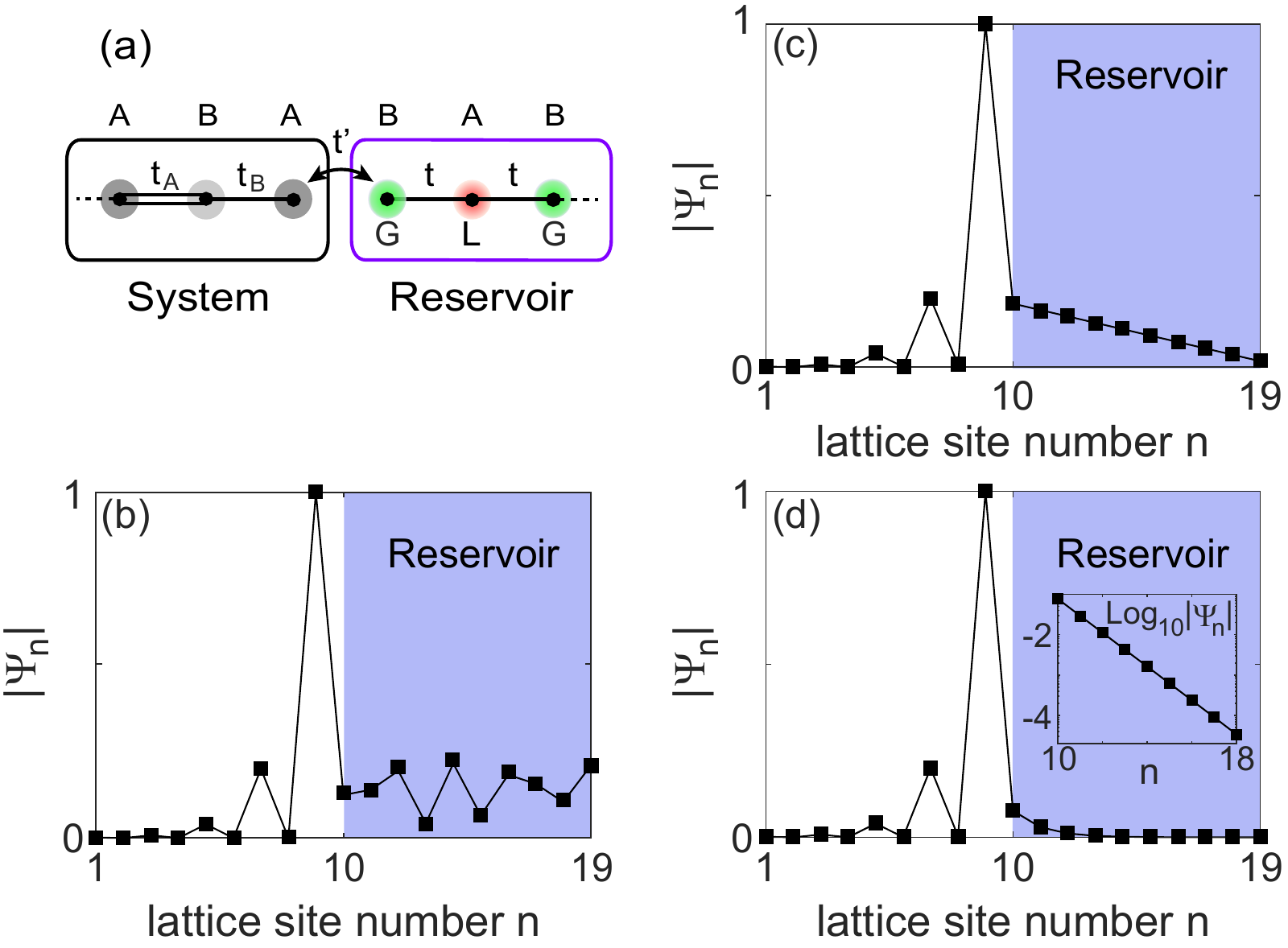}
\caption{\textbf{Model and non-Hermitian zero modes in different localization regimes.} (a) Schematic of an SSH chain on the left coupled weakly to a non-Hermitian reservoir on the right. $G$ and $L$ indicate gain and loss sites. (b--d) Spatial profile of the zero mode with $\im{\omega_\mu}\approx0$ when $\gamma/t=0.5,2,3$. Squares and solid lines show results from exact numerics and first-order perturbation theory, respectively. $t_A/t=1$ and $t_B/t=t'/t=0.2$ are used, and we find $\omega_\mu/t=0.0251i$, $0.0356i$, and $0.0147i$ in these three figures. Inset in (d): logscale plot of the exponential tail in the reservoir.}\label{fig:schematic}
\end{figure}

\noindent \textbf{Model.} A symmetry-protected zero mode in a Hermitian system is often found when the spectrum of the system is symmetric, satisfying $\omega_\mu = -\omega_\nu$. A zero mode then emerges when the two mode indices $\mu,\nu$ are identical, resulting in $\omega_\mu=0$. Such a symmetric spectrum can be the result of chiral symmetry or particle-hole symmetry \cite{Hasan,Qi,Alicea,Beenakker,Sarma_RMP}. 
The spectrum of a non-Hermitian system, on the other hand, is generally complex, and hence it is necessary to restore the complex conjugation in the manifestation of particle-hole symmetry, i.e., $\omega_\mu = -\omega_\nu^*$, which are symmetric about the imaginary axis. This relation leads to the definition of a non-Hermitian zero mode, i.e., $\re{\omega_\mu}=0$ when $\mu=\nu$, which was shown to be a general property of many gain and loss modulated lattices \cite{zeromodeLaser} (see also Supplementary Note 1).

\begin{figure}[t]
\centering
\includegraphics[width=\linewidth]{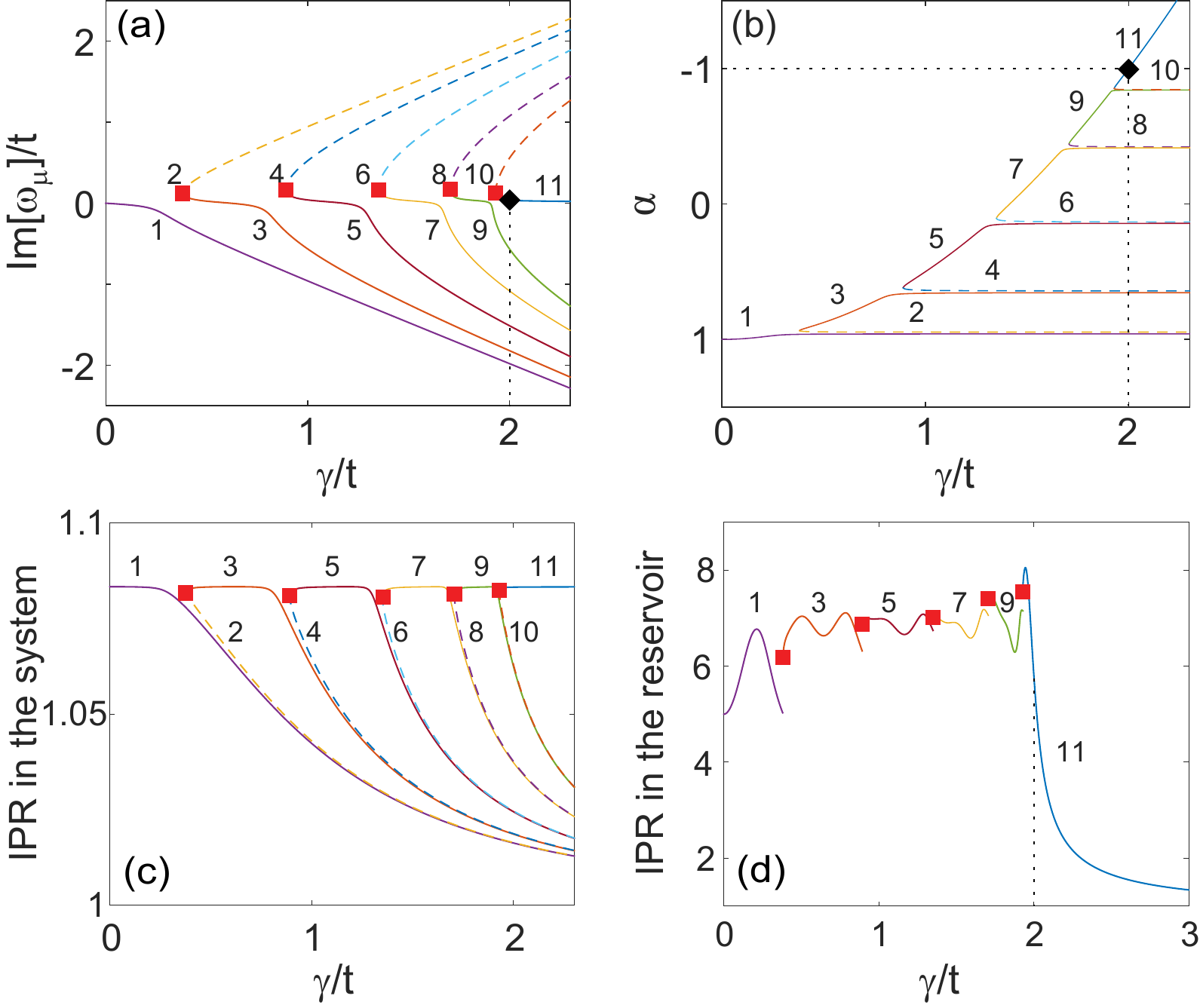}
\caption{\textbf{Properties of non-Hermitian zero modes.} (a) Non-Hermitian zero modes (solid and dashed lines) as a function of $\gamma$. Red squares mark EPs, and the same in (c) and (d). (b) The critical quantity $\alpha$ in these non-Hermitian zero modes. The dotted lines mark $\alpha\approx1$ at $\gamma=2t$. (c) IPR of the non-Hermitian zero modes in the system. (d) IPR of the odd-numbered zero modes in the reservoir. Dotted line marks $\gamma=2t$.}\label{fig:transition}
\end{figure}

The ``system'' we consider below is a Hermitian SSH chain [left half in Fig.~\ref{fig:schematic}(a)] with real on-site potential $\omega_0$:
\be
H_S  = \sum_n \omega_0 |n\rangle \langle n|  + \left(\,t_n|n+1\rangle \langle n|  + h.c.\,\right). \label{eq:Hs}
\ee
Here the nearest-neighbor (NN) coupling $t_n$ is given by $t_A$ ($t_B$) when $n$ is odd (even). This system is coupled from the right via the NN coupling $t'$ to a non-Hermitian ``reservoir'' [right half in Fig.~\ref{fig:schematic}(a)]:
\be
H_R  = \sum_n \omega_n |n\rangle \langle n|  + \left(\,t|n+1\rangle \langle n|  + h.c.\,\right). \label{eq:Hr}
\ee
$t$ is its NN coupling, and $\omega_n$ is the complex on-site potential, given by $\omega_A=\omega_0-i\gamma$ ($\omega_B=\omega_0+i\gamma$) when $n$ is odd (even)  representing loss (gain) with $\gamma>0$. Note that a tight-binding model for a finite lattice does not require an explicit boundary condition; by definition, this lattice is isolated from its environment except for the assumed gain and loss (e.g., $\gamma$ in our case). Due to the challenge of fabricating and coupling many nearly identical elements in micro- and nano-photonics, here we consider a small number of lattice sites, i.e., 9 in the system and 10 in the reservoir. Note that the entire lattice does not have parity-time symmetry \cite{NPreview}, as reflected by the absence of up and down symmetry in Fig.~\ref{fig:transition}(a). This property is independent of the number of lattice sites.

\noindent \textbf{Non-Hermitian zero modes and linear localization.} Henceforth we take $\omega_0=0$ as the zero energy. It is well known that topological and (chiral) symmetry protected zero mode(s) exist in the SSH model \cite{SSH}, and in our case it is exponentially localized on the right side of the system when $t_A>t_B$. When coupled to the non-Hermitian reservoir, chiral symmetry of the SSH chain is broken, but non-Hermitian particle-hole symmetry still holds for the combined lattice \cite{zeromodeLaser}. It pins the original Hermitian zero mode on the imaginary axis [mode 1 in Fig.~\ref{fig:transition}(a)]. As the gain and loss coefficient $\gamma$ increases, more non-Hermitian zero modes are generated (modes 2 to 11): A pair of non-zero modes satisfying $\omega_\mu=-\omega_\nu^*$ can coalesce at their EP on the imaginary axis, then repel each other along the imaginary axis (see Supplementary Figure 1). This process takes place five times in this combined lattice, giving rise to five EPs [see red squares in Fig.~\ref{fig:transition}(a)]. 

While all these eleven non-Hermitian zero modes are similarly localized in the system half [as indicated by the inverse participation ratio $\text{IPR}={(\sum_i |\Psi_i|^2 )^2}/{\sum_i |\Psi_i|^4}$ in the range [1,1.1]; see Fig.~\ref{fig:transition}(c)], there is a clear lineage across the odd-numbered ones with an $\text{IPR}\approx1.08$. This phenomenon is a non-Hermitian analogy of avoided crossings in Hermitian system but now on the imaginary axis [see Fig.~\ref{fig:transition}(a)], through which a non-Hermitian zero mode passes its identity to the next mode, including its $\im{\omega_\mu}$ and IPR in the system. 

By following this lineage, we find that the tails of the odd-numbered zero modes inside the non-Hermitian reservoir experience a transition from delocalized [Fig.~\ref{fig:schematic}(b)] to localized [Fig.~\ref{fig:schematic}(d)] when $\gamma$ increases, also verified by the IPR calculated in the reservoir [Fig.~\ref{fig:transition}(d)]. During this transition, we observe a linear tail in mode 11 at $\gamma\approx2t$ [Fig.~\ref{fig:schematic}(c)], where not just $\re{\omega_\mu}=0$ (which defines a non-Hermitian zero mode) but also $\im{\omega_\mu}$ is  approximately zero ($\omega_\mu/t=0.0356i$). Its tail in the reservoir has an $R^2$ value of 0.9997, close to being perfectly linear. This qualitative change of the wave function is confirmed using a first-order perturbation theory (see solid lines in Fig.~\ref{fig:schematic}), where we treat the weak coupling $t'$ as the small perturbation parameter (see Method). In addition, we also find that the correction to the zero-mode energy is zero, which is qualitatively correct and shines the light on our later analysis using the recurrence relation (\ref{eq:iter}).

\noindent \textbf{Origin of linear localization.} To probe whether this linear tail is related to an EP of a Hamiltonian, we first note that the critical value $\gamma=2t$ does not produce an EP of the combined lattice; the latter, as Fig.~\ref{fig:transition}(a) shows, all take place in $\gamma<2t$, and hence their corresponding wave functions are delocalized in the reservoir (see Supplementary Figure 1). Furthermore, if we treat the reservoir as periodic, its Bloch Hamiltonian
\be
H_k = \begin{pmatrix}
i\gamma & t(1+e^{ik}) \\
t(1+e^{-ik}) & -i\gamma 
\end{pmatrix}
\ee
has two bands $\omega_\pm(k)=\pm\sqrt{2t^2(1+\cos k)-\gamma^2}$, where $k\in[-\pi,\pi]$ is the momentum. These two bands exhibit an EP ring in the $k$-$\gamma$ parameter space (see Supplementary Figure 2), at $\omega=0$ and given by $k=\pm \cos^{-1} [\gamma^2/(2t^2)-1]$. At $\gamma=2t$, this EP ring gives an EP at $k=0$, where $H_k$ becomes the Hamiltonian of the classical parity-time symmetric dimer \cite{NPreview}; its coalesced eigenstate $\psi=[i,1]^T$ with $\omega_\pm(0)=0$ or the corresponding Jordan vector $J=[0,i]^T$ (defined by $H_kJ=\psi$) does not imply a linear tail, either. Finally, the eigenstates of the finite-sized reservoir do not display a linear tail at this critical value of $\gamma$. The corresponding eigenvalues are not EPs either, with the ones closest to $0$ at $\omega_\mu/t=\pm i0.563$; their wave functions are best approximated by sine functions of a half period (see Supplementary Figure 3).  

Having excluded the cause of linear localization from an EP of a Hamiltonian, either of the combined lattice or the reservoir itself, we focus on the recurrence relation (\ref{eq:iter}). To derive this relation, we first rewrite equation~(\ref{eq:Hr}) in the non-Hermitian reservoir as
\be
\Psi_n = i\frac{\kappa_{A,B}}{t} \Psi_{n-1} - \Psi_{n-2}\label{eq:iter2}
\ee
for an eigenstate, where $\kappa_{A,B}\equiv-i(\omega_\mu-\omega_{A,B})$. For a non-Hermitian zero mode we have $\re{\omega_\mu}=0$, and hence $\kappa_{A,B}=\im{\omega_\mu}\pm\gamma$ are real and they represent the effective loss and gain coefficients on the $A$ and $B$ sublattices in the reservoir. Next, by applying equation~(\ref{eq:iter2}) to five consecutive lattice sites and eliminating $\Psi_{n-1},\Psi_{n-3}$, we arrive at equation~(\ref{eq:iter}) with
\be
\alpha = -\left(1+\frac{\kappa_A\kappa_B}{2t^2}\right)=-\left(1+\frac{\im{\omega_\mu}^2-\gamma^2}{2t^2}\right)\in \mathbb{R}.
\ee
Besides the trivial solution $\Psi_n=0$ on all lattice sites, the recurrence relation (\ref{eq:iter}) also permits the Bloch solution
\be
\Psi_n =\beta_1 b_+^m + \beta_2 b_-^m,\label{eq:Bloch}
\ee
where $\beta_{1,2}$ are two constants and the integer $m$ is the index on a sublattice, i.e., $m=(n+1)/2$ on the $A$ sublattice (where $n$ is odd) and $m=n/2$ on the $B$ sublattice (where $n$ is even). $b_\pm$ are the two roots of the characteristic equation
\be
b^2-2\alpha b + 1 = 0\label{eq:character}
\ee
of equation~(\ref{eq:iter}), or more explicitly, $b_\pm=\alpha\pm i\sqrt{1-\alpha^2}\equiv e^{\pm i\theta}$ where $\theta=\cos^{-1}\alpha$.

When $|\alpha|<1$, $\theta$ is real and we have $\Psi_n = (\beta_1+\beta_2)\cos m\theta + i(\beta_1-\beta_2)\sin m\theta$, 
leading to a $|\Psi_n|$ oscillating between the upper bound $|\beta_1|+|\beta_2|$ and the lower bound $||\beta_1|-|\beta_2||$ as $n$ (and $m$) varies. In other words, the non-Hermitian zero modes are delocalized in the reservoir, as we have seen in Fig.~\ref{fig:schematic}(b) where $\alpha = -0.875$. When $|\alpha|>1$, $\theta$ is imaginary and $\Psi_n$ is the linear superposition of two exponential functions, with one decaying toward the left and the other toward the right. The latter then leads to the localization in the reservoir, as we have shown in Fig.~\ref{fig:schematic}(d) where $\alpha = 3.5$.

The critical behavior takes place at $\alpha=\pm1$, where the characteristic equation (\ref{eq:character}) has a single root $b=\alpha$ of multiplicity 2. Besides the (now) single solution given by equation (\ref{eq:Bloch}), i.e., $\Psi_n\propto \alpha^m$, there also exists a non-Bloch solution proportional to $m\alpha^m$ \cite{MathBook}, with which we write the general solution as
\be
\Psi_n =\alpha^m (\rho_1+\rho_2m),\quad\rho_{1,2}=const. \label{eq:sol2}
\ee
The term linear in $m$ can then potentially lead to linear localization if $\rho_2\neq0$. We note though $\alpha=-1$ does not cause linear localization, because it requires one or both of $\kappa_{A,B}$ to be zero, and equation~(\ref{eq:iter2}) tells us immediately that $|\Psi_n|$ is a constant on at least one sublattice and hence delocalized. Therefore, the linear localization shown in Fig.~\ref{fig:schematic}(c) is the result of $\alpha = 1$, which is achieved in mode 11 at $\gamma\approx2t$ with $\im{\omega_\mu}\approx0$ [Fig.~\ref{fig:transition}(b)]. 

To determine $\rho_{1,2}$ in equation~(\ref{eq:sol2}), i.e., the offset and slope of the linear tail, we study the recurrence relation at both ends of the reservoir for $\gamma=2t$ and $\im{\omega_\mu}=0$. For convenience, we relabel the rightmost site as 1 and the leftmost site in the reservoir as $N_R$ (10 in our case). By revisiting equation~(\ref{eq:iter2}) for the last three sites on the right (i.e., $\Psi_3 = -2i\Psi_2 - \Psi_1$) and that for just the last two (i.e., $\Psi_2=2i\Psi_1$), we find $\rho_1=0$ (i.e., no offset of the linear tail) and 
\be
\Psi_n=
\Bigl\{\;
    \begin{matrix}
    \rho_2 n/2, &(n\; \text{odd}) \\
    i\rho_2 n/2, &(n\; \text{even})
\end{matrix}\label{eq:linear}
\ee
inside the reservoir. Similarly, by studying the equation for $\Psi_{N_R}=i\rho_2 N_R/2$ that couples to the rightmost site in the system with a normalized wave function $\Psi_0=e^{i\phi}\,(\phi\in\mathbb{R})$,
i.e.,
\be
i\partial_t \Psi_{N_R} = 0 = i\gamma (i\rho_2 N_R/2) + t'e^{i\phi} + t \rho_2 (N_R-1)/2,\nonumber
\ee
we find $\phi=0$ or $\pi$ and 
\be
\rho_2 = \pm\frac{2}{1+N_R}\frac{t'}{t}. \label{eq:slope}
\ee
In other words, the slope of the linear tail is proportional to the relative strength of the system-reservoir coupling ($t'$) over that in the reservoir ($t$). It is also inversely proportional to the size of the reservoir in the large $N_R$ limit, where we find that $|\Psi_{N_R}|=|\rho_2|N_R/2$ is fixed at $t'/t$, and the linear tail simply stretches as the reservoir becomes longer until the effect of the EP between modes 10 and 11 in Fig.~\ref{fig:transition} becomes significant (see Supplementary Figure 4 and the discussion at the end of the next subsection).
\\

\begin{figure}[b]
\centering
\includegraphics[width=\linewidth]{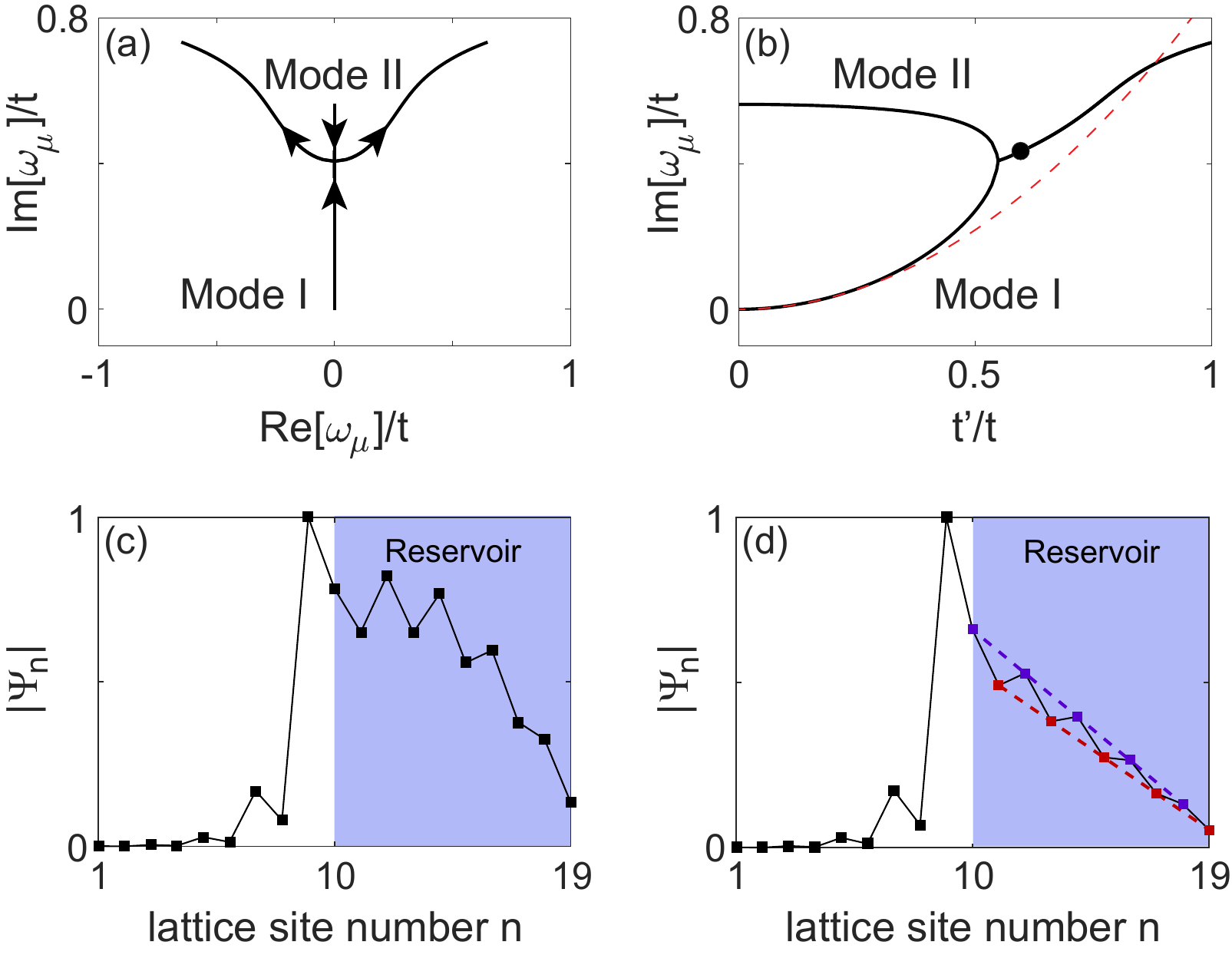}
\caption{\textbf{Strong coupling between system and reservoir and zigzag linear localization.} (a) Evolution of two modes near $\omega=0$  as $t'$ increases with a fixed $\gamma=2t$ in the reservoir. (b) Their imaginary parts as a function of $t'$. Red dashed line shows the result of the second-order perturbation theory (see Methods), which agrees well with the numerical result up to $t'/t\approx0.4$. Black dot marks $t'/t=0.6$, where modes I and II have an identical spatial profile shown in (c). (d) Zigzag linear localization of the zero mode at $t'=0.6t$, with $\gamma=2.036t$ and $\omega_\mu/t=0.3823i$. Dashed lines show the linear behavior of $|\Psi_n|$ on the two sublattices in the reservoir. Other parameters are the same as in Fig.~\ref{fig:schematic}(c). }\label{fig:zigzag}
\end{figure}

\noindent \textbf{Strong coupling regime.} The results above for $\alpha=1$, again, are only valid if we have also $\im{\omega_\mu}=0$ at $\gamma=2t$ for the non-Hermitian zero mode. As we have mentioned, the prediction $\re{\omega_\mu}=\im{\omega_\mu}=0$ from the perturbation theory holds only when the system-reservoir coupling $t'$ is weak. When $t'$ is strong and comparable to $t$, then $\im{\omega_\mu}\approx0$ breaks down (and even $\re{\omega_\mu}=0$ breaks down), and linear localization disappears. 

As an example, we show in Figs.~\ref{fig:zigzag}(a,b) the evolution of two modes near $\omega=0$ as a function of $t'$ with a fixed $\gamma=2t$. When $t'=0$, i.e., the system and the reservoir are decoupled, mode I is the original Hermitian zero mode of the SSH chain localized at site 9, and mode II is the non-Hermitian zero mode of the reservoir mentioned before in Supplementary Figure 3, with its wave function given approximately by sine functions of a half period. They move along the imaginary axis and coalesce at an EP when $t'/t\approx0.547$, after which they move off the imaginary axis and are no longer non-Hermitian zero modes. Fig.~\ref{fig:zigzag}(c) shows the identical spatial profile of modes I and II at $t'/t=0.6$, and clearly they do not display a linear tail in the reservoir. 

We note that unlike the case of varying $\gamma$ shown in Fig.~\ref{fig:transition}(a), there is no succession of EPs on the imaginary axis as $t'$ increases. In fact, the EP mentioned above is the only one that exists in $t'>0$ (the situation in $t'<0$ is the same as the sign of $t'$ does not change the eigenvalues of the modes). This observation can be attributed to the fact that at $t'=0$, all modes of the system are on the real axis while those of the reservoir are on the imaginary axis. They have minute couplings except between modes I and II discussed above, and this pair are simply given by $\omega_\mu\approx\pm t'+i\gamma/2$ in the large $t'$ limit.

Nevertheless, we can still find a proper $\gamma$ that realizes $\alpha=1$ but away from $\gamma=2t$ with a finite $\im{\omega_\mu}$. In this case, equation~(\ref{eq:sol2}) only indicates that the wave function of the non-Hermitian zero mode is linear on \textit{each} of the two sublattices (given by even and odd numbered sites in the reservoir), because the recurrence condition at the right boundary that leads to equation~(\ref{eq:linear}) is now different. An example is given in Fig.~\ref{fig:zigzag}(d), where $\alpha=1$ is achieved at $t'=0.6t$ with $\gamma=2.036t$. Such zigzag linear localization can also be found when the reservoir itself is an non-Hermitian SSH chain (see Supplementary Figure 5).

We also note that although $t'=0.2t$ is weak compared to other relevant energy scales with the reservoir shown in Fig.~\ref{fig:schematic}, it becomes strong effectively near an EP. This is similar to what happens near an avoided crossing in quantum mechanics, where the wave functions of two coupled modes become highly hybridized. The same situation near an EP can also destroy linear localization. For example, as the reservoir becomes much longer, the EP between modes 10 and 11 shown in Fig.~\ref{fig:transition} approaches $\gamma=2t$, and the system becomes highly sensitive to the value of $\gamma$. At exactly $\gamma=2t$, the spatial profiles of modes 10 and 11 are both hybridization of the original edge state in the system and the sine-like mode in the reservoir (see Supplemental Figure 4). The approximation $\im{\omega_\mu}\approx0$, while still true (e.g., $\im{\omega_\mu}=0.0415$ with $N_R=134$), no longer justifies equation~(\ref{eq:linear}) due to the enhanced sensitivity near the EP. However, if we slightly change $\gamma/t$ to $3.78\times10^{-4}+2$, we find $\alpha=1$ and achieve zig-zag linear localization (see Supplementary Figure 4). In this case, the two linear tails on the two sublattices are well aligned with each other, giving rise to an $R^2$ value of 0.9996.


\begin{figure}[t]
\centering
\includegraphics[width=\linewidth]{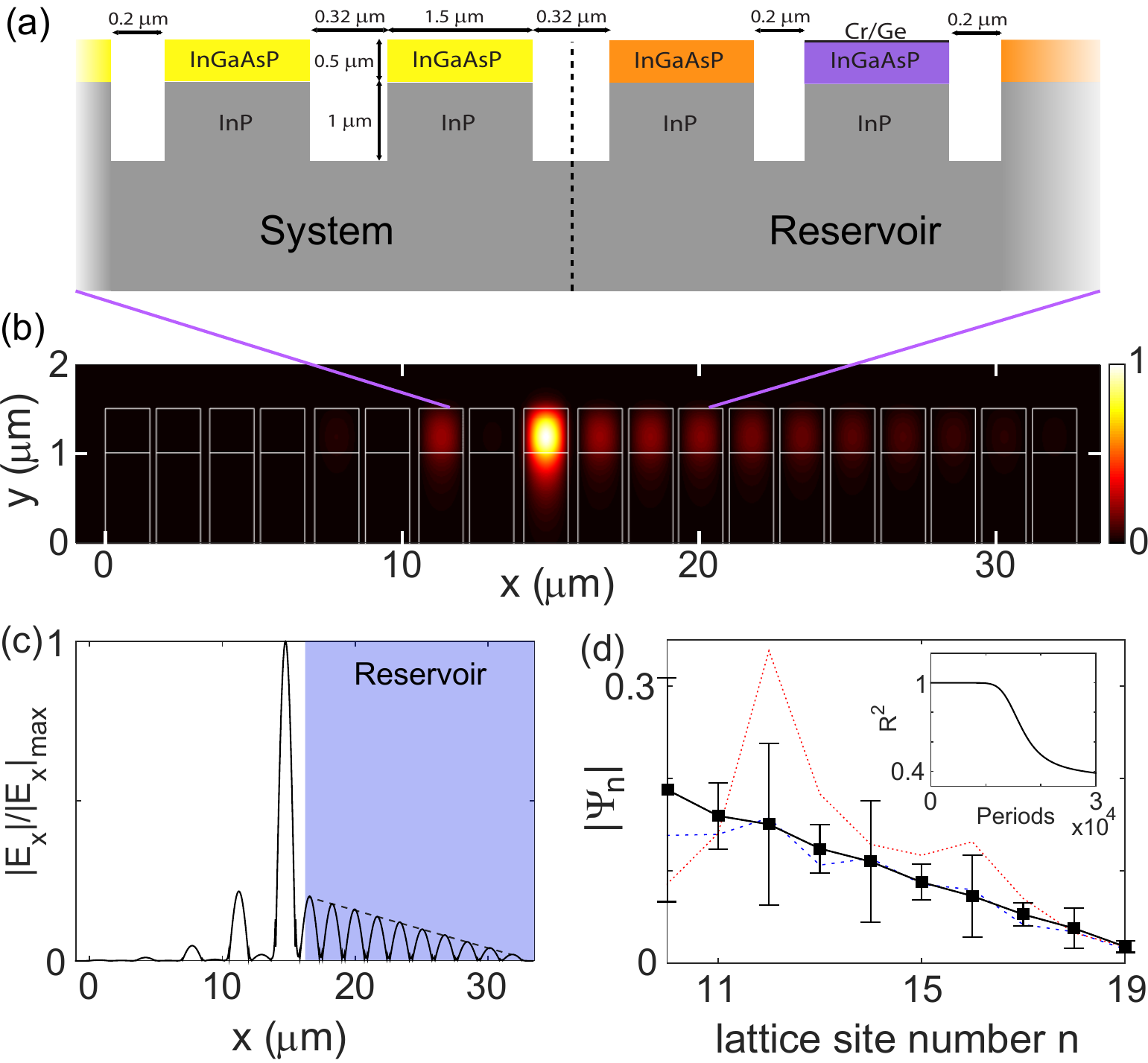}
\caption{\textbf{Photonics simulations of linear localization.} (a) Schematic of coupled InP waveguides in both the system and the reservoir. Color schemes for different InGaAsP regimes are: pumped to the transparent threshold (yellow), gain (orange), and loss with the Cr/Ge layer (purple). (b) False color plot of the $x$-component of the electric field $|E_x|$ in a linearly localized non-Hermitian zero mode. (c) $|E_x|$ at $y\approx1.2\mu{m}$ in (b). Dashed linear line is shown as a comparison. (d) Ensemble average of $|\Psi_n|$ in the reservoir. Error bars show the standard deviations. Dashed and dotted lines show a typical initial state and a final state after $10^4$ periods. Inset: $R^2$ value of the ensemble average.}\label{fig:waveguide}
\end{figure}

\noindent \textbf{Photonic simulations.} The tight-binding model, similar to those given by equations~(\ref{eq:Hs}) and (\ref{eq:Hr}) [see also Methods], has been employed successfully to capture various phenomena in coupled waveguide systems. Some references include, for example, the experimental demonstration of parity-time symmetric photonics \cite{Guo}, the observation of supersymmetric mode converters \cite{Heinrich}, and the creation of a non-Hermitian zero-mode \cite{Pan}.

To showcase the validity of the tight-binding model in our study, i.e., linear localization can be demonstrated on a realistic photonic platform, we consider in Fig.~\ref{fig:waveguide}(a) evanescently coupled InP waveguides \cite{Miao,Wong}.
The electric field propagating along the waveguide direction ($z$) can be written as $\vec{\Psi}(x,y,z)=\vec{E}(x,y)e^{-i\beta z}$ \cite{Klaiman}, where $z$ and the propagation constant $\beta$ play the roles of time and frequency in our analytical model studied above. We convert $\beta$ to the effective index using $n_\text{eff}=\beta\lambda/2\pi$, where the wavelength $\lambda=1.55 \mu$m. This allows us to express the couplings as dimensionless quantities (i.e., in the unit of the effective index). Using finite-difference-time-domain simulations of the vectorial Maxwell's equations \cite{meep}, we find that $t_A=t=5.37\times10^{-5}$ and $t_B=t'\approx0.24t$ for an air gap of $0.2\mu$m and $0.32\mu$m between neighboring waveguides, respectively. 
 
Optical gain in the reservoir is modeled by pumping the InGaAsP quantum wells, while loss is provided by a thin Cr/Ge double layer \cite{Feng} placed on top of alternate waveguides that also blocks the pump light. We denote the corresponding imaginary parts of the intrinsic indices in the gain and loss waveguides by $\pm n''$, which is reduced by a factor of $0.86$ in the effective index. By increasing $n''$ to approximately $1.24\times10^{-4}$, we bring the effective gain and loss in the reservoir to twice the value of $t$, and we indeed 
identify the target mode displaying linear localization in the reservoir [Figs.~\ref{fig:waveguide}(b,c)], confirming the prediction of our analytical model. Its effective index is found to be $n_\text{eff}=n_0+2.24i\times10^{-6}$, where $n_0=3.255$ is the effective index in a single waveguide. In other words, this mode is indeed a non-Hermitian zero mode and it amplifies slowly down the waveguides. 

Although this zero mode does not have the highest gain [which is the case also in the tight-binding model; see mode 11 in Fig.~\ref{fig:transition}(a)], we find it to be robust in the presence of noise and can be observed after an ensemble average. Using the extracted parameters from the simulations, we prepare $10^3$ initial states with noises, i.e., with the wave function of this zero mode multiplied by $e^{0.1s}$ in the reservoir, where $s$ varies from site to site and follows the standard normal distribution. We then let them evolve up to $10^4$ periods (about 4.8 mm down the waveguides). Even though the tail of the final state in each run can be highly nonlinear in space, their ensemble average maintains a good linear profile, with an $R^2$ value of 0.9966 [Fig.~\ref{fig:waveguide}(d)].

\section*{Discussion}

In summary, we have shown that the wave function of a non-Hermitian zero mode can display unusual linear localization in a weakly coupled non-Hermitian reservoir, which takes place during the transition between extended and exponentially localized regimes. Although we have considered a gain and loss modulated reservoir, linear localization can also be observed in an all-passive lattice by the inclusion of an additional on-site loss that exceeds $\gamma=2t$ at all lattice sites, including both the system and the reservoir. The system coupled to the non-Hermitian reservoir is not limited to 1D; higher dimensions systems with zero modes display linear localization as well under the same conditions (see Supplementary Figure 6). The finding of this exotic behavior for topology and symmetry protected zero modes urges us to rethink localization in non-Hermitian systems. This is especially so in the emerging interdisciplinary field that bridges two of the most energized fields in physics, namely the studies of topological phases of matter and non-Hermitian photonics based on novel symmetries, especially with the surging interest in the non-Hermitian skin effect.



We thank Vadim Oganesyan, Sarang Gopalakrishnan, Bo Zhen, Steven Johnson, and Aaron Welters for helpful discussions. This project is supported by the NSF under Grant No. PHY-1847240. A brief summary of an early version was included as a proceeding in the submission of an unpresented poster \cite{poster}.

\section*{Methods}

\noindent \textbf{Hamiltonian of the entire lattice.} We have specified the Hamiltonian $H_S$ of the Hermitian ``system" and the effective Hamiltonian $H_R$ of the non-Hermitian ``reservoir" using equations (\ref{eq:Hs}) and (\ref{eq:Hr}) in their second quantization forms. The corresponding matrix forms are given by
\be
H_S =
\begin{pmatrix}
\ddots & \ddots   &          &          &          \\
\ddots & \omega_0 & t_B      &          &          \\
       &  t_B     & \omega_0 & t_A &    &          \\
       &          & t_A      & \omega_0 & t_B      \\
       &          &          & t_B      & \omega_0
\end{pmatrix},
\ee
\be
H_R =
\begin{pmatrix}
\omega_0 +i\gamma   & t                   &                    &             \\
t                   & \omega_0 - i\gamma  & t                  &             \\
                    & t                   & \omega_0 + i\gamma & \ddots      \\
                    &                     & \ddots             & \ddots
\end{pmatrix},
\ee
and the effective Hamiltonian of the entire lattice can then be expressed as:
\be
H =
\begin{pmatrix} [c c c | c c c]
\ddots & \ddots   &          &          &                   \\
\ddots & \omega_0 & t_B      &          &                   \\
       &  t_B     & \omega_0 & t' &    &                    \\
       \hline
       &          & t'      & \omega_0+i\gamma & t          &          \\
       &          &          & t      & \omega_0-i\gamma    & \ddots    \\
       &          &          &        & \ddots              & \ddots
\end{pmatrix}.\label{eq:totalH}
\ee

\vspace{0.5cm}
\noindent \textbf{Perturbation theory.} In Fig.~\ref{fig:schematic} we have verified the results of the tight-binding model using a first-order perturbation theory. Here we first outline this approach. We treat the weak coupling $t'$ between the system and the reservoir as the small perturbation parameter, and we denote $H_0=H_S+H_R(\gamma)$ as the unperturbed (and uncoupled) Hamiltonian of the system ($H_S$) and the reservoir ($H_R$). In the coupled system the Hamiltonian becomes
\be
H=H_0+t'H'\quad(t'\neq0),
\ee
where $H'$ contains only two nonzero elements in its matrix form, which are 1's coupling the rightmost lattice site in the system and the leftmost site in the reservoir [see equation~(\ref{eq:totalH})]. The right and left eigenstates of $H_0$ are denoted by $|\psi_\mu^{(0)}\rangle$ and $\langle\tilde{\psi}_\mu^{(0)}|$, and they satisfy the biorthogonal relation
\be
\langle\tilde{\psi}_\mu^{(0)}|\psi_\nu^{(0)}\rangle = \delta_{\mu\nu} \label{eq:biortho}
\ee
away from an EP of $H_0$. Because $H_0$ is symmetric, i.e. $H_0=H_0^T$, we find $\tilde{\psi}_\mu^{(0)} = {\psi}_\mu^{(0)}$ for all modes.

Note that because there is no coupling between the system and the reservoir in $H_0$, $|\psi_\mu^{(0)}\rangle, \langle\tilde{\psi}_\mu^{(0)}|$ are both zero either in the system or the reservoir. There are then two consequences. First, the biorthogonal relation above is nontrivial only when both modes $\mu$ and $\nu$ belong to the system or the reservoir simultaneously. Second, an EP of $H_0$ must be an EP of either $H_S$ or $H_R$. Clearly, there is no EP in the (Hermitian) SSH part of the Hamiltonian, and we show in Supplementary Figure 3 that there is no EP in a finite-sized reservoir either when linear localization takes place (i.e., at $\gamma\approx2t$). Therefore, we can apply equation~(\ref{eq:biortho}) in the perturbation theory to study linear localization.

Similar to the perturbation theory in (Hermitian) quantum mechanics, the first-order correction to the energy is given by
\be
\omega_\mu^{(1)}=t'\langle\tilde{\psi}_\mu^{(0)}|H'|\psi_\mu^{(0)}\rangle\equiv t'H'_{\mu\mu}.
\ee
Due to the spatial properties of $|\psi_\mu^{(0)}\rangle, \langle\tilde{\psi}_\mu^{(0)}|$ mentioned above and the structure of $H'$, it is straightforward to show that $H'_{\mu\nu}$ is zero if both $|\psi_{\mu}^{(0)}\rangle, \langle\tilde{\psi}_\nu^{(0)}|$ are in the system or reservoir. As a result, $H'_{\mu\mu}$ (and hence $\omega_\mu^{(1)}$) are zero for all eigenstates of $H$, i.e., the first-order correction to the energy of the original SSH zero mode is zero, which we have used in the derivation of equation~(\ref{eq:linear}) in the main text.

In the meanwhile, the first-order correction to the wave function is given by
\be
|\psi_\mu^{(1)}\rangle = t'\sum_{\nu\neq\mu}\frac{H'_{\nu\mu}}{\omega_\mu^{(0)}-\omega_\nu^{(0)}}|\psi_\nu^{(0)}\rangle,
\ee
which is plotted in Fig.~1 of the main text for different values of $\gamma$. We note that this correction does not change the zero-mode wave function in the system; it only changes the latter in the reservoir, again due to the property of $H'_{\mu\nu}$ mentioned above.

We have shown the result for the second-order correction to the energy in Fig.~\ref{fig:zigzag}(b). It displays the characteristic quadratic dependence on the perturbation parameter (i.e., $t'$), which is given by
\be
\omega_\mu^{(2)}={t'}^2 \sum_{\nu\neq\mu}\frac{{H'_{\mu\nu}}^2}{\omega_\mu^{(0)}-\omega_\nu^{(0)}}.
\ee
Note that we do not take the absolute value of the numerator, which differs from the (Hermitian) perturbation theory in quantum mechanics. This feature is required to perturbe the energy to become complex, even with real-valued $\omega_{\mu,\nu}^{(0)}$ as we have here.\\

\renewcommand\figurename{\hspace{-1mm}}    
\renewcommand\thefigure{\textbf{Supplementary Figure \arabic{figure}}}
\setcounter{figure}{0}
\renewcommand\theequation{S\arabic{equation}}
\setcounter{equation}{0}

\noindent \textbf{Supplementary Note 1: Non-Hermitian particle-hole symmetry}
\vspace{2mm}

\begin{figure}[h]
\centering
\includegraphics[width=\linewidth]{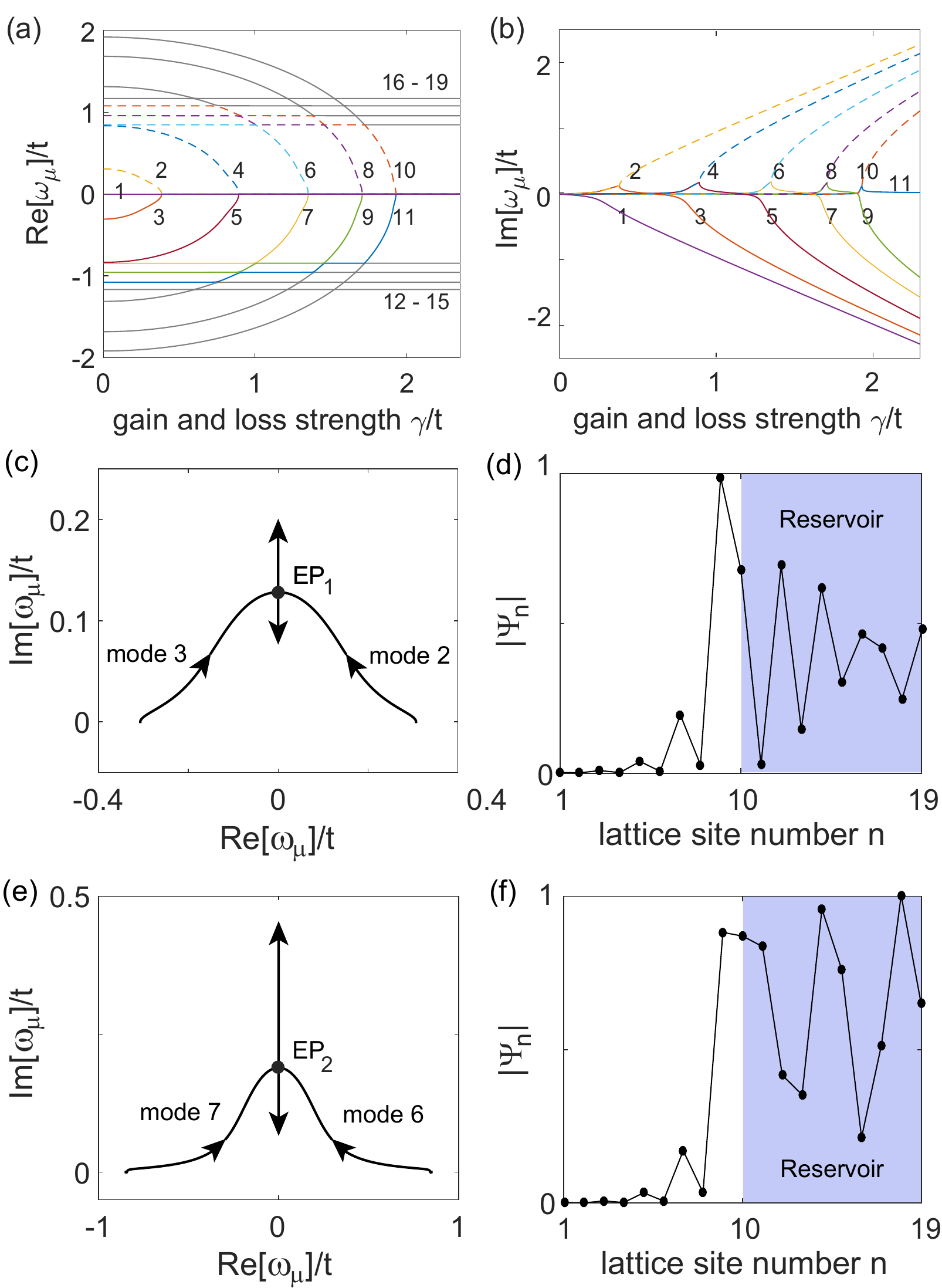}
\caption{(a,b) Real and imaginary parts of the spectrum of our combined lattice. (c) Trajectories of modes 2 and 3 when the gain and loss strength $\gamma$ in the reservoir is increased from 0 to $0.4t$. $t$ is the uniform nearest-neighboring coupling in the reservoir. These two modes coalesce at an exceptional point (EP$_1$) when $\gamma=0.384t$. Their identical spatial profile at EP$_1$ is shown in (b). (c,d) Same as (a,b) but for modes 6 and 7. $\gamma\in[0,1.4t]$ and the exceptional point (EP$_2$) occurs at $\gamma=1.351t$.}\label{fig:wf_EP}
\end{figure}

As we will show in \ref{fig:wf_EP}, all the EPs in our combined chain are on the imaginary axis, and they are determined by the particle-hole symmetry \cite{zeromodeLaser,NHFlatband_PRL,Ge2018} instead of parity-time symmetry. Below we briefly review this symmetry and its consequences.

First, we require two sublattices (denoted by ``A'' and ``B'') with no detuning, which means that (1) the matrix representation of the Hamiltonian has zero diagonal elements; and that (2) the couplings only take place between two lattice sites on two different sublattices. These requirements lead to chiral or sublattice symmetry in a Hermitian system. 
As a result, we can write the tight-binding Hamiltonian and its eigenvector in the following way:
\be
H_0 =
\begin{pmatrix}
0 & T_1 \\
T_2 & 0
\end{pmatrix}\in \mathbb{R},
\quad
\Psi = 
\begin{pmatrix}
\Psi^{(A)}_1\\
\vdots\\
\Psi^{(A)}_{N_A}\\
\Psi^{(B)}_1\\
\vdots\\
\Psi^{(B)}_{N_B}\\
\end{pmatrix}\label{eq:H_chiral}
\ee
where $T_1$ is a $N_A\times N_B$ matrix while $T_2$ is a $N_B\times N_A$ matrix, containing all the couplings. $N_A$ and $N_B$ are the numbers of lattice sites on the A and B sublattices. The wave function is represented by a column vector, with its values on each sublattice grouped together (i.e., $\Psi^{(A)}_1,\ldots,\Psi^{(A)}_{N_A}$ and $\Psi^{(B)}_1,\ldots,\Psi^{(B)}_{N_B}$). In the case of our reservoir with alternate gain and loss, its effective Hamiltonian can be written in the form of Eq.~(\ref{eq:H_chiral}) above with $\Psi^{(A)}_1,\ldots,\Psi^{(A)}_{N_A}$ being the wave function on the loss lattice sites (``sublattice A'') and $\Psi^{(B)}_1,\ldots,\Psi^{(B)}_{N_B}$ on the gain lattice sites (``sublattice B''). Such a partition is possible even when gain and loss are random or mixed (see below).

Secondly, we require the couplings to be real. They do not need to be symmetric and can be completely random. 

Finally, we require a block diagonal matrix $H'$ with only imaginary elements. Note that they can be completely random, with the positive (negative) values representing optical gain and loss. With this in mind, we can write down the full Hamiltonian
\be
H = H_0 + H' = 
\begin{pmatrix}
D_1 & T_1 \\
T_2 & D_2
\end{pmatrix}
\ee
and a symmetry operator
\be
C = 
\begin{pmatrix}
I_1 &  \\
 & I_2
\end{pmatrix}
\ee
where $I_1$ ($I_2$) is the $N_A\times N_A$ ($N_B \times N_B$) identity matrix and $D_1$ ($D_2$) is an imaginary $N_A\times N_A$ ($N_B \times N_B$) random matrix. It is then straightforward to show that
\be
\{CK,H\} = CKH+HCK=0\label{eq:NHPH}
\ee
using the fact that $D_{1,2}$ are imaginary. Here $K$ is the complex conjugation. When $D_{1,2}$ are diagonal, they represent optical gain and loss at each lattice site (representing a waveguide or resonator), and we still have two sublattices in the original sense. When $D_{1,2}$ are not diagonal, this sense is lost since we now have imaginary couplings between sites in the same sublattice as well. 

The anticommutation relation (\ref{eq:NHPH}) defines the non-Hermitian particle-hole symmetry here, and consequently, the eigenvalues of $H$ satisfy $\omega_\mu=-\omega_\nu^*$, where $\mu,\nu$ are different in general but can be the same. When $\mu,\nu$ are different, it can be shown that $\Psi_\mu=CK\Psi_\nu$, or in other words, the non-Hermitian particle-hole symmetry is spontaneously broken in the two corresponding eigenstates $\Psi_{\mu,\nu}$. Such a pair of eigenvalues are symmetric about the imaginary axis, and they can and do coalesce on the imaginary axis as exemplified in \ref{fig:wf_EP}. When $\mu,\nu$ are the same (and $\re{\omega_\mu}=0$), we have $\Psi_\mu=CK\Psi_\mu$ instead, which restores the non-Hermitian particle-hole symmetry.
\\

\noindent \textbf{Supplementary Figure 1: EPs of the combined lattice}
\vspace{2mm}

In Figure~2(a) of the main text, we have shown the trajectories of all the zero modes in our coupled lattice, with 9 (10) lattice sites in the system (reservoir) and an increasing gain and loss parameter $\gamma$. In \ref{fig:wf_EP}(b) below, we also show the imaginary parts of mode 2 to 11 before they become zero modes. \ref{fig:wf_EP}(a) shows their real parts, together with the non-zero modes 12 to 19. 
The up-and-down symmetry in this panel is a result of non-Hermitian particle-hole (NHPH) symmetry we have elucidated in \textbf{Supplementary Note 1}. \ref{fig:wf_EP}(c),(e) show two examples of how two non-zero modes become non-Hermitian zero modes after coalescing at their EP on the imaginary axis.

Using Figure~2(a) in the main text, we have emphasized that linear localization does not occur at the EPs of our combined lattice. Here we exemplify the wave functions at these EPs in \ref{fig:wf_EP}(d),(f), which are extended in the reservoir. The reason that there are five such EPs is because the whole lattice, including both the system and the reservoir, has 19 modes as determined by the number of lattice sites (i.e., 19). By varying $\gamma$ in the reservoir, only those modes extending well into the reservoir change significantly, and there are eleven of them (Modes 1–11). They include the ten modes originated from the reservoir, now coupled weakly to the system, and the edge mode of the system, now coupled weakly to the reservoir. These eleven modes satisfy the non-Hermitian particle-hole symmetry, and they form five ``doublets,'' satisfying $\omega_\mu=-\omega_\nu^*$, and a ``singlet,'' i.e., a non-Hermitian zero mode with $\re{\omega_\mu}=0$. Each doublet coalesce on the imaginary axis at the corresponding $\gamma$, giving rise to five EPs in total.
\\

\noindent \textbf{Supplementary Figure 2: EPs in the band structure of the reservoir}
\vspace{2mm}

\begin{figure}[t]
\centering
\includegraphics[width=\linewidth]{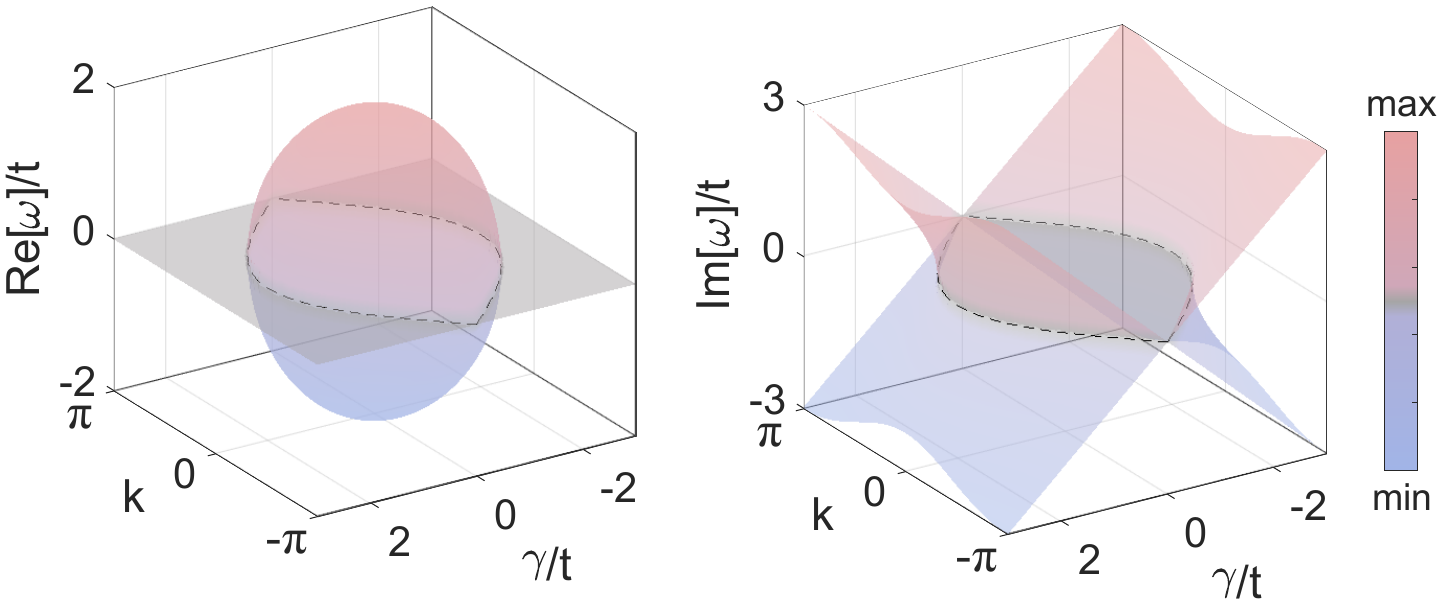}
\caption{Exceptional ring in the first Brillouin zone of the reservoir, plotted as a function of $k$ and $\gamma$.}\label{fig:EPring}
\end{figure}

We mentioned in the main text that an EP ring exists in the band structure of the reservoir, when the reservoir is treated as periodic and infinite in size. In \ref{fig:EPring} we show this EP ring, both in the real and imaginary parts of the band energy. Note that a negative $\gamma$ represents the time-reversed reservoir, with gain and loss exchanged. Because EP(s) occur for every value of $\gamma\in(0,2t]$ in the first Brillouin zone whereas linear localization takes place only at $\gamma=2t$ (and $-2t$, its time-reversed partner), they do not have a causal relationship. The system is Hermitian when $\gamma=0$, and hence this point is excluded from the EP ring. 
\\

\noindent \textbf{Supplementary Figure 3: Eigenstates of the reservoir at $\gamma=2t$}

The reservoir alone has NHPH symmetry (independent of its size) and PT symmetry (when there are an even number of lattice sites).
The eigenvalues $\omega_\mu$'s of its Hamiltonian still follow closely the result of the periodic case, i.e., $\omega_\pm(k)=\pm i\sqrt{\gamma^2-2t^2(1+\cos k)}$ where $k\in[-\pi,\pi]$. In other words,  
$\omega_\mu$'s are either on the real axis or imaginary axis, with a maximum $|\re{\omega_\mu}| = \sqrt{4t^2-\gamma^2}$ and a maximum $|\im{\omega_\mu}| = \gamma$ when $\gamma\in[0,2t]$. 

\begin{figure}[h]
\centering
\includegraphics[width=\linewidth]{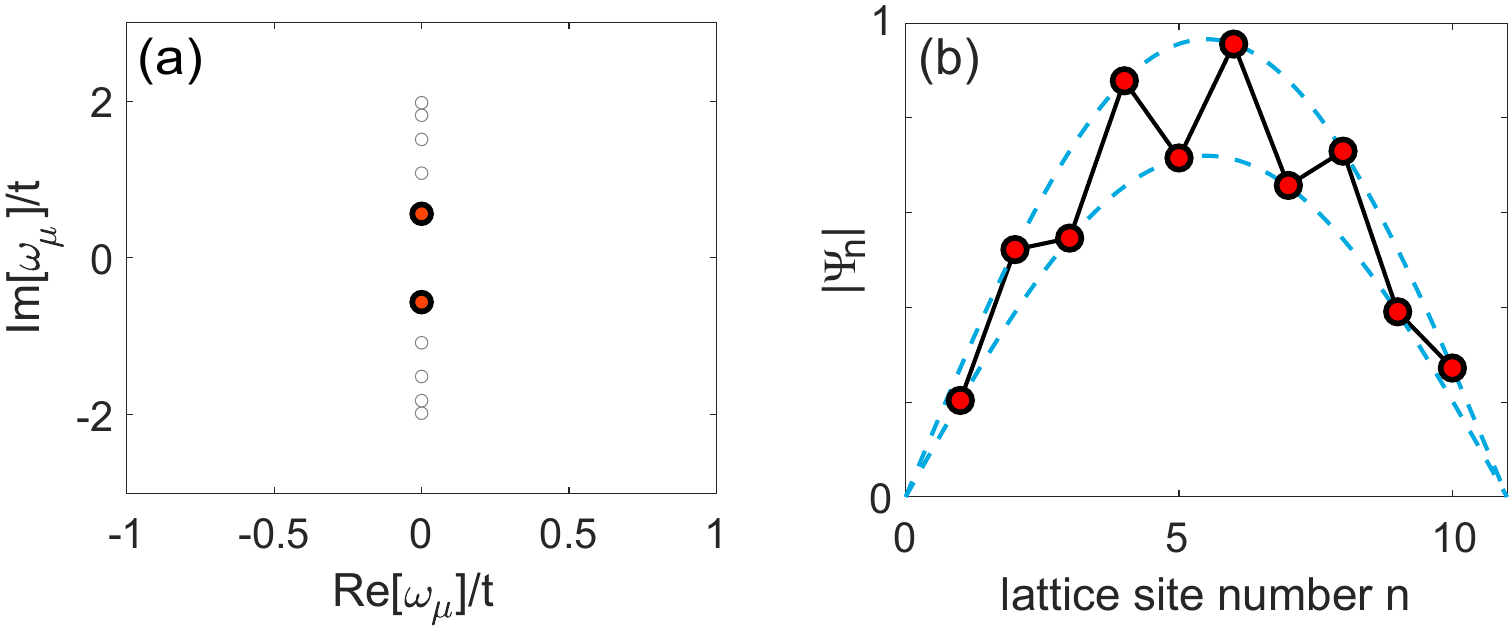}
\caption{(a) Eigenvalues of the reservoir Hamiltonian with $\gamma=2t$. The spatial profiles of the two closest to the origin (filled circles) are mirror images of each other, and (b) shows the one with a positive $\im{\omega_\mu}$, where the dashed lines are two sine functions of a half period, one for each sublattice. 
}\label{fig:reservoir}
\end{figure}

For example, at $\gamma=2t$, all these eigenvalues of the reservoir itself are on the imaginary axis with $|\im{\omega_\mu}|<2t$ [\ref{fig:reservoir}(a)]. The wave functions of the ones with the smallest absolute value (i.e., $\omega_\mu/t=\pm i0.563$) are PT-symmetric, and their spatial profiles $|\Psi|$ are mirror images of each other, which can be captured by a single sine function for each sublattice [\ref{fig:reservoir}(b)].

\vspace{2cm}

\noindent \textbf{Supplementary Figure 4: Stretched linear tail in longer non-Hermitian reservoirs}
\vspace{2mm}

\begin{figure}[h]
\centering
\includegraphics[width=\linewidth]{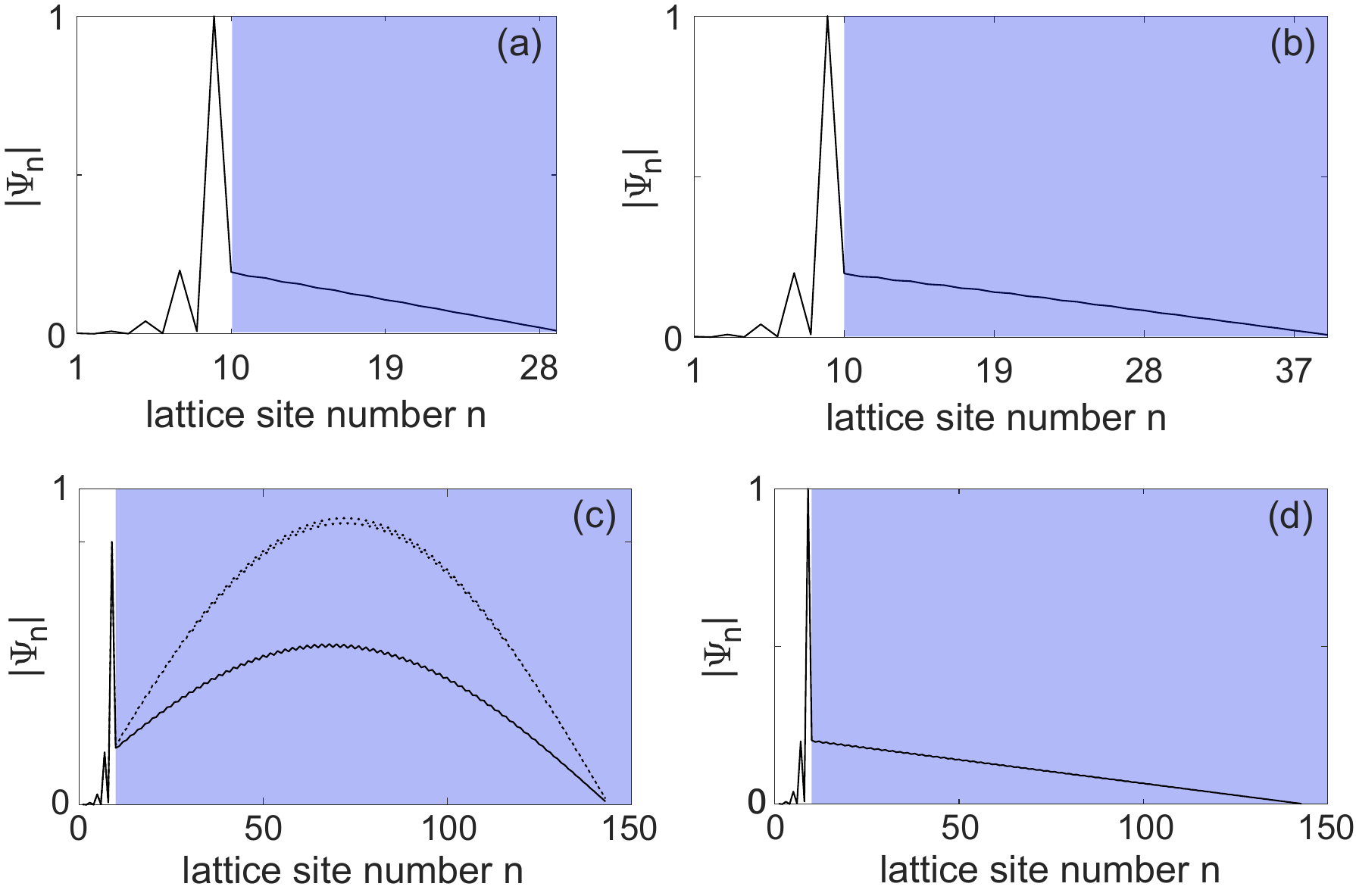}
\caption{(a,b) Stretched linear tail in longer reservoirs with 20 and 30 sites in the reservoir, respectively. Other parameters are the same as in Figure~1(c) of the main text. (c) Disappearance of linear tail near an EP, when the reservoir has 132 sites. Dotted and solid lines show modes 10 and 11, respectively. (d) Restored zig-zag localization from (c) with $\gamma/t=2+3.78\times10^{-4}$. }\label{fig:longerReservoir}
\end{figure}

We show in \ref{fig:longerReservoir} the scaling of linear localization when the reservoir is extended. It corroborates our proof in the main text that the linear tail is simply stretched without losing its linearity. The $R^2$ value is 0.9996 and 0.9992 for these two cases shown in panels (a) and (b), respectively. 

This behavior breaks down when $\gamma=2t$ is near an EP (i.e., the one between modes 10 and 11 in Fig.~2(a) of the main text) where high-order effects modify linear localiztion. Panel (c) shows the case where there are 132 lattice sites in the reservoir, and the frequencies of modes 10 and 11 are very close ($\omega_\mu=0.0415i,0.0437i$) to their EP. Their spatial profiles are similar, given by the hybridization of the original edge state in the system and the sine-like mode in the reservoir at $\gamma=2t$. By slightly varying $\gamma/t$ to $2+3.78\times10^{-4}$, we find $\alpha=1$ for mode 11 and achieve a zig-zag linear localization that is almost aligned on the two sublattices [panel (d)], as mentioned in the main text. The dramatic change of the wave function inside the reservoir indicates the strong sensitivity near the EP.
\\

\noindent \textbf{Supplementary Figure 5: Zigzag linear localization in a non-Hermitian SSH reservoir}
\vspace{2mm}

\begin{figure}[t]
\centering
\includegraphics[width=\linewidth]{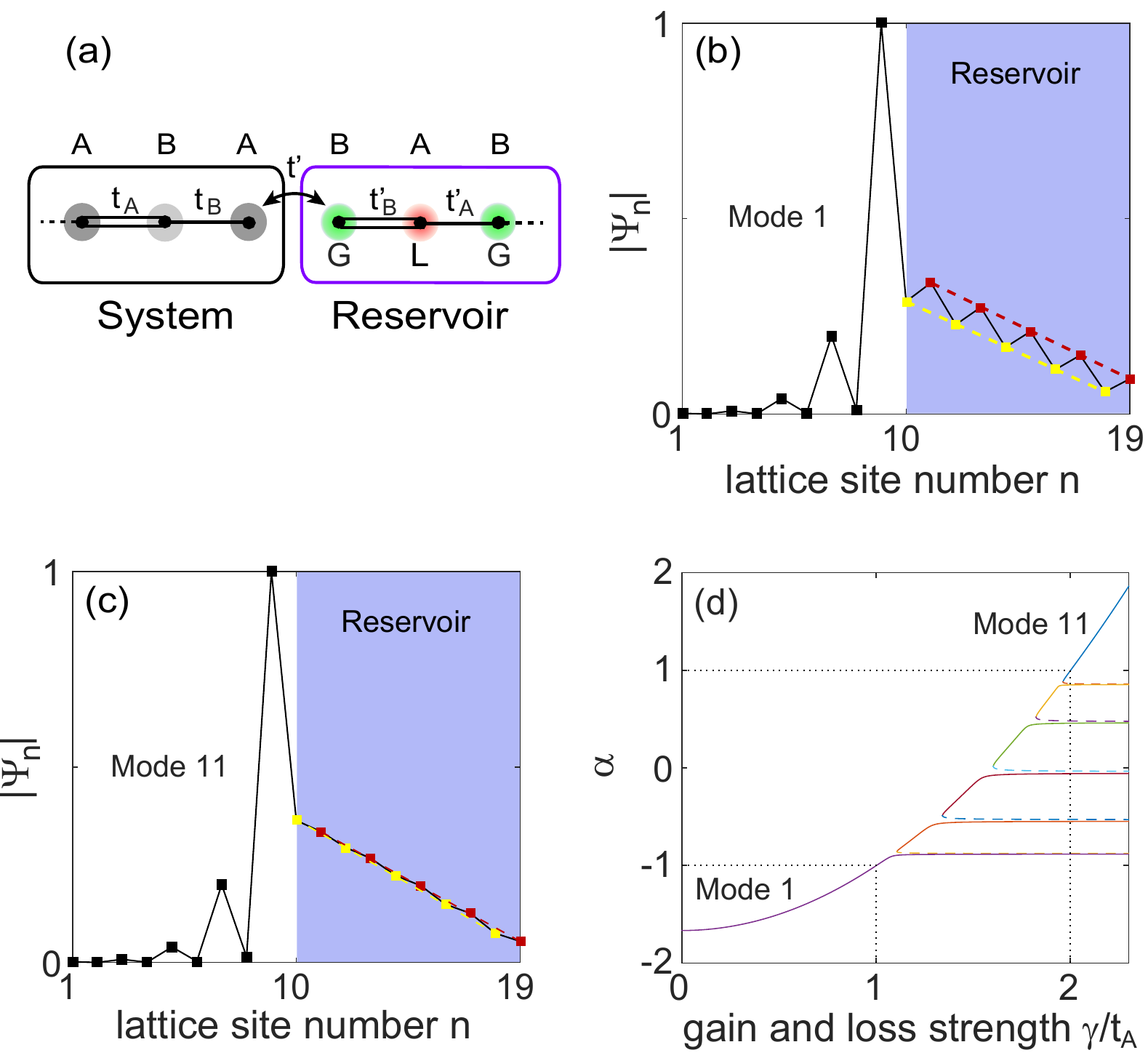}
\caption{Zigzag linear localization of the non-Hermitian zero mode near $\im{\omega_\mu}=0$ at $\gamma/t_A=1$ (b) and 2 (c). The system and the reservoir are the same as in Figure~1 of the main text, except that the uniform coupling $t$ in the reservoir is replaced by alternate $t'_{A,B}/t_A=1.5,0.5$ shown schematically in (a). (d) $\alpha$ as a function of the gain and loss strength. }\label{fig:SM1}
\end{figure}

We show here zigzag linear localization in a non-Hermitian SSH reservoir [see \ref{fig:SM1}(a)], which features alternate gain and loss as well as coupling strengths $t'_{A,B}>0$. The linear recurrent relation given by Eq.~(1) in the main text, i.e.,
\be
\Psi_n = 2\alpha \Psi_{n-2} - \Psi_{n-4}
\ee
still holds with a generalized
\begin{align}
\alpha &= -\frac{1}{2}\left(\frac{t'_A}{t'_B} + \frac{t'_B}{t'_A} + \frac{\kappa_A\kappa_B}{t'_A t'_B}\right) \nonumber \\
&= -\frac{1}{2}\left(\frac{t'_A}{t'_B} + \frac{t'_B}{t'_A} + \frac{\im{\omega_\mu}^2-\gamma^2}{t'_A t'_B}\right)\in\mathbb{R}. \label{eq:alpha2}
\end{align}
Again with $\im{\omega_\mu}\approx0$, zigzag linear localization is achieved at both $\gamma \approx t'_A+t'_B$ and $|t'_A-t'_B|$ [see \ref{fig:SM1}(b),(c)], corresponding to $\alpha\approx\pm 1$ [see \ref{fig:SM1}(d)]. 
\\

\noindent \textbf{Supplementary Figure 6: Linear localization with a 2D system}
\vspace{2mm}

\begin{figure}[htb]
\centering
\includegraphics[width=\linewidth]{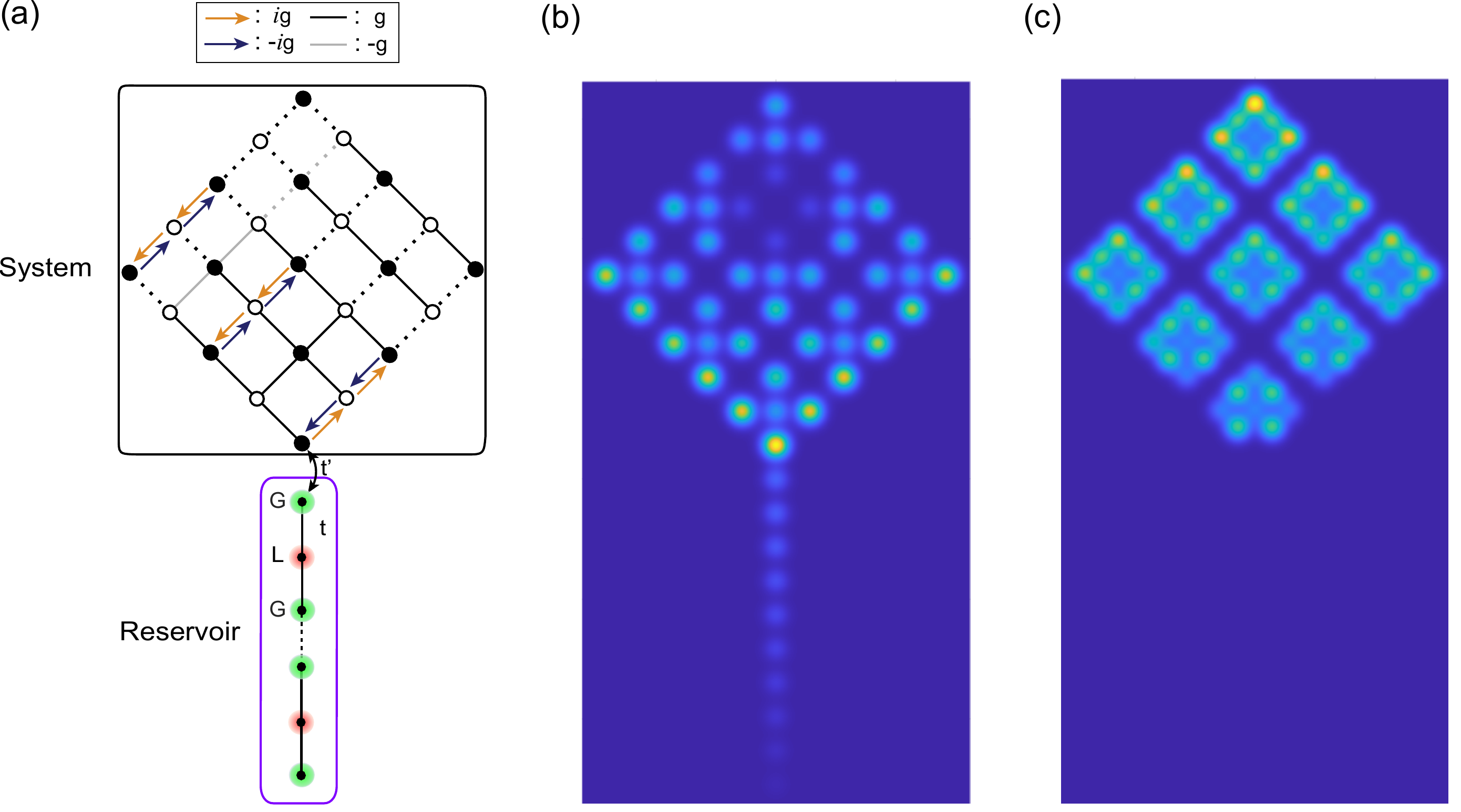}
\caption{(a) Schematic showing the topological ``system'' and the non-Hermitian ``reservoir''. (b,c) Two zero modes that couples and does not couple to the reseroir, respectively.}\label{fig:topo}
\end{figure}

As mentioned in Discussion of the main text, the dimension of the system coupled to the non-Hermitian reservoir is not limited to 1D. Here we show linear localization when the system is a two-dimensional topological photonic lattice shown in \ref{fig:topo}. It has the same coupling $g$ in the $-45^\circ$ direction, and the couplings in the $+45^\circ$ direction can be asymmetric (but still Hermitian), with the same amplitude and a phase that changes by $\pi/2$ successively in the $-45^\circ$ direction. This configuration can be realized using spatially displaced ring couplers \cite{Hafezi1,Hafezi2}, and it leads to a synthetic gauge field for photons, with each of the smallest plaquettes pierced by a flux of $\pi/2$. Non-Hermitian extensions of this model have led to topological insulator lasers \cite{
topolaser1,topolaser2}, on-demand routing along gain and loss boundaries \cite{toporouting}, and the demonstration of a hidden conservation law due to pseudo-chirality \cite{pseudoChirality}.

When an $11\times11$ lattice is considered, there are three zero modes in this system. We then couple one corner of it to the same non-Hermitian reservoir used in Fig.~1 of the main text, with 10 sites of alternate gain and loss at $\gamma=2t$, $g/t=1$, and $t'/t=0.2$. Two of the original zero modes become decoupled from the reservoir [see \ref{fig:topo}(c), for example], without changing their frequencies. The one that does couple to the reservoir displays linear localization in the latter [\ref{fig:topo}(b)], with an $R^2\approx0.999997$. Its frequency is shifted up slightly along the imaginary axis to $\omega_\mu/t\approx3.2\times10^{-3}i$, which again satisfies the criterion $\omega_\mu\approx0$ as in the 1D case shown in the main text.     

\end{document}